\documentclass[preprint,aps,prd,groupedaddress,showpacs,nofootinbib]{revtex4-1}%
\usepackage{graphicx}
\usepackage{amsmath}
\usepackage{amssymb}
\usepackage{bm}
\usepackage{epsfig}
\usepackage{dcolumn}
\usepackage{slashed}
\usepackage{empheq}

\newcommand{\bea}{\begin{eqnarray}}
\newcommand{\eea}{\end{eqnarray}}
\newcommand{\nn}{\nonumber}

\usepackage[colorlinks,linkcolor=blue,anchorcolor=blue,citecolor=blue,urlcolor=blue]{hyperref}

\begin{document}

\setlength\baselineskip{17pt}

\title{ Two-loop $N$-jettiness soft function for $pp \to 2j$ production }

\author{Shan Jin}
\email{shanjin@mail.bnu.edu.cn}
\affiliation{Physics Department, Beijing Normal University, Beijing 100875, China}

\author{Xiaohui Liu}
\email{xiliu@bnu.edu.cn}
\affiliation{Physics Department, Beijing Normal University, Beijing 100875, China}
\affiliation{Center of Advanced Quantum Studies, Beijing Normal University, Beijing 100875, China}

\date{\today}

\begin{abstract}
We calculate the two-loop soft function for $pp \to jj$, $e^+e^- \to 4j$ or $e p \to 3j$ when the $N$-jettiness observable is measured. 
The result presented here makes up the necessary piece for realizing the full next-to-next-to-leading order predictions for these
processes using the $N$-jettiness subtraction scheme. 
\end{abstract}

\maketitle

\section{Introduction}
 The accumulation of the data set and the continuously increasing precision of the experimental analyses bring
the physics program at the Large Hadron Collider (LHC) and other experiments the ability 
to search for new physics from the small deviations of the data from the Standard Model predictions. In order to fulfill such searches, the precision of the theoretical predictions from Quantum Chromodynamics (QCD) has to match the small experimental errors to reliably interpret the collider data. Currently, the only theoretical tool, with {\textit no} known substitutes, enables us to make predictions for collider phenomenologies out of the first QCD principle is the perturbative calculation based on the systematic expansion of the strong coupling constant $\alpha_s$ or $\alpha_s L$ if large logarithmic correction $L$ is present. In recent years, tremendous efforts have been made in realizing the theoretical predictions for the physical processes at the colliders with a full control of the final state kinematics beyond the next-to-leading order (NLO) accuracy.  On the fixed order side, significant progress has been made in the past few years, which includes achieving the 
fully differential gluon-gluon fusion Higgs ($H$) production~\cite{Cieri:2018oms, Dulat:2018bfe} and the single jet production in deep-inelastic scattering~\cite{Currie:2018fgr} at next-to-next-to-next-to-leading order (N$^3$LO) in $\alpha_s$ and the 
next-to-next-to-leading order (NNLO)  calculations of the single inclusive jet or dijet productions~\cite{Currie:2017eqf}, $V/H + j$ processes~\cite{Boughezal:2015dva,Ridder:2015dxa,Boughezal:2015ded,Gehrmann-DeRidder:2017mvr,Boughezal:2015dra, Chen:2014gva,Boughezal:2015aha}, Higgs production through the vector boson fusion~\cite{Cacciari:2015jma,Cruz-Martinez:2018rod}, single top~\cite{Brucherseifer:2014ama, Berger:2016oht} and $t{\bar t}$ productions~\cite{Czakon:2015owf, Catani:2019iny} at the LHC, the jet productions in deep-inelastic scattering~\cite{Abelof:2016pby, Currie:2017tpe}, and the heavy flavor production in deep-inelastic neutrino scattering~\cite{Berger:2016inr}. 

The rapid growth in the higher order predictions benefit from either the new higher loop calculations achieved or the theoretical schemes developed for dealing with the singularities in the real emissions. Recent achievements of the analytic two-loop five-point amplitude~\cite{Abreu:2018aqd, Chicherin:2018yne, Chicherin:2018old} highlight the current status of the loop calculations for the collider processes, and meanwhile new ideas are keeping on popping out~\cite{Borowka:2018dsa,Liu:2017jxz,Liu:2018dmc,Xu:2018eos}. To handle the infrared (IR) singularities in the real corrections, various schemes have been proposed and successfully applied to the LHC phenomenology, such as the local subtraction schemes including the antenna subtraction~\cite{GehrmannDeRidder:2005cm}, the sector improved residue subtraction~\cite{Czakon:2010td} and the nested soft-collinear subtraction~\cite{Caola:2017dug}, as well as the physical observable based global schemes like the $q_T$-subtraction~\cite{Catani:2007vq}, inclusive jet mass subtraction~\cite{Gao:2012ja} and the $N$-jettiness subtraction~\cite{Boughezal:2015dva,Gaunt:2015pea}. Although the global schemes are always criticized for the numerical instability, they have realized many interesting higher order predictions for the physical processes at the hadron colliders including the numerically most challenging Higgs production at N$^3$LO~\cite{Cieri:2018oms}.

Among the local schemes, $N$-jettiness scheme is the known scheme being able to handle the IR singularities for generic jet processes at the hadron colliders. The scheme utilizes a threshold cut-off $\tau_{\rm cut}$ in the $N$-jettiness event shape $\tau$~\cite{Stewart:2010tn} to distinguish between the NNLO $N$ jet (when $\tau < \tau_{\rm cut}$) and the NLO $N+1$ jet configurations ($\tau > \tau_{\rm cut}$). Below the $\tau_{\rm cut}$, a Soft-Collinear-Effective-Theory~\cite{Bauer:2000ew,Bauer:2000yr,Bauer:2001ct,Bauer:2001yt,Bauer:2002nz} approach based on the factorization theorem~\cite{Stewart:2010tn}
\bea
\mathrm{d}  \sigma(\tau_{\rm cut}) = 
 B \, B \, {\rm Tr}[S \cdot H]  \, \prod_{i = 1} J_i + {\cal O}\left(\tau_{\rm cut} \right) \,, 
\eea
is used to approximate the full QCD contribution. Here $B$ stands for the beam function due to the initial state energetic radiations while $J$'s are the jet functions for the final state collinear radiations. The hard function $H$ encodes the hard loop information. $S$ is the soft function for soft radiations. The trace is over the relevant color space. All the ingredients in the factorization theorem are universal except for the hard function $H$ which depends on the specific process under consideration.
The difference between the factorization theorem and the full QCD is suppressed by power corrections in $\tau_{\rm cut}$, which have been studied extensively recently for the $0$-jettiness case~\cite{Moult:2016fqy,Boughezal:2016zws,Moult:2017jsg,Boughezal:2018mvf,Ebert:2018lzn}. 
The universal beam and the jet functions are all known analytically to two-loops~\cite{Gaunt:2014xga,Gaunt:2014cfa,Becher:2006qw,Becher:2010pd} and recently some of them are even calculated at three-loop order~\cite{Bruser:2018rad, Banerjee:2018ozf, Melnikov:2018jxb}. As for the quark beam function, its NNLO longitudinal polarized counterpart is also known in the literature~\cite{Boughezal:2017tdd}. The calculation of the soft function is complicated by its explicit dependence on the full $N$-jettiness measure, therefore its analytic NNLO calculation is by far out of reach and people turn to numerical solutions. A generic framework was proposed in~\cite{Boughezal:2015eha} for calculating such soft functions with arbitrary numbers of jets and the $1$-jettiness soft functions in $pp$-collision were calculated using this scheme or similar approach for both the massless jet~\cite{Boughezal:2015eha, Campbell:2017hsw} and the massive heavy flavor productions~\cite{Li:2016tvb, Li:2018tsq}. However, at NNLO, the most general calculations can only be covered when one considers the $N$-jettiness soft function with four external hard legs, which is missing in the original calculations. In this manuscript, we present such calculation by extending the previously developed numerical method and show sufficient calculating details. We noticed a recent conference proceeding~\cite{Bell:2018mkk} on the $2$-jettiness soft function using \texttt{pySecDec}. Our work uses a different and independent 
 approach which can serve as a cross check of the calculations.  

The rest of this work is organized as follows. We discuss the numerical setups and the phase space parameterization in detail in Section~\ref{sec:setup}. In Section~\ref{sec:num}, we present our numerical results for the $2$-jettiness observable in hardon-hadron collisions. Last we conclude in Section~\ref{sec:conclude}

\section{Computational Setup}\label{sec:setup}
In this section, we summarize the setups used in our calculation. We follow the general numerical framework 
proposed in~\cite{Boughezal:2015eha}, and show its capability in dealing with $N$-jettiness with four hard reference vectors in $pp\to jj$, $ep \to 3j$ and $0 \to 4j$. We note that the calculations 
with four reference vectors go through all possible configurations in calculating any $N$-jettiness soft functions at NNLO.  
We clarify some of the computational details of this framework. 

For simplicity, we stick to the hadronic $N$-jettiness definition which reads
\bea
\tau = \sum_i {\rm min}\left(
n_1 \cdot q_i\,, n_2 \cdot q_i \,,
\dots \,, 
n_N \cdot q_i
\right) \,,
\eea
where $n_i$'s are the $N$ light-like hard reference vectors along the initial state hadronic beams or the
 final state jet axes and $q_i$'s are the four momentum
of the soft radiations. 
We note that here all the hard vectors $n_i$'s are physical objects and are defined in $4$-dimension while the soft momentum $q_i$'s can potentially be un-resolvable and are therefore $d$-dimensional 
quantities, with $d =  4-2\epsilon$ in the dimensional regularization. 
The soft function with $n$ real soft emissions then can be calculated by performing  
the phase space integration 
\bea
S(\tau) = 
\prod^n_i
\int  \, 
\frac{ \mathrm{d}^d  q_i }{(2\pi)^{d-1} }  \delta^+(q_i^2)  \, 
{\cal S}(\{q_i\}) \, \Theta_{N,i}(\{q_i\}) \,, 
\eea
where we have introduced the $N$-jettiness measurement function
\bea
\Theta_{N,i} (\{q_i\}) = \sum_{j=1}^N
\delta(\tau - 
\sum_i^n
n_{j_i} \cdot q_i) \prod_{k_i \ne j_i }^{N-1} \theta( n_{k_i} \cdot q_i - n_{j_i} \cdot q_i) \,,
\eea
and ${\cal S}$ is the soft current matrix element for the soft radiations which, up to NNLO, has been known for a long time~\cite{Catani:1999ss, Catani:2000pi}. 
For $N$-jettiness with $3$ reference vectors , to NNLO, the soft current can be expressed in terms of a sum of dipole contributions, while starting from
four reference vectors, like the case we study here, new triple-pole structure involving 3 eikonal lines arises which is induced by the one-loop soft current.
The triple-pole contribution, which will be calculated later in this work, is the only remaining piece for calculating the $N$-jettiness soft function with arbitrary
external eikonal lines at ${\cal O}(\alpha_s^2)$. For completeness, we summarize the explicit form of the soft current ${\cal S}$ to ${\cal O}(\alpha_s^2)$ in the Appendix~\ref{app:soft-current}.

In order to compute the soft function numerically, the key step is to find a suitable momentum parameterization to allow us to 
write the phase space integration in the form of  
\bea\label{eq:Laurent}
\int_0^1 \prod_i \mathrm{d} x_i \, x_i^{-1 - a_i \epsilon}  \, F(\{x_i \}) 
 = \int_0^1 \prod_i \mathrm{d} x_i \, \left[  - \frac{1}{a_i \, \epsilon } \delta(x_i) +  \left( \frac{1}{x_i} \right)_+ 
+  \dots 
  \right]  \, F(\{x_i \}) 
\,, 
\eea
which is ready for the Laurent expansion 
so as for us to identify all the $\epsilon$-poles and evaluate the coefficients of the poles numerically.  
We thus now turn to the discussion of the phase space parameterization for 
single and double real emissions with four reference directions, separately. 
We note that though we are focusing on the four reference vector case, the parameterization is general enough to be applicable directly to any $N$-jettiness computation at the NNLO. In all cases below, we utilize the Sudakov decomposition to write an arbitrary light-like four vector $k^\mu$ with respect to 
 two of the reference vectors $n_i$ and $n_j$ as
 \bea
 k^\mu = \frac{n_j^\mu}{n_{ij}} k^+ +  \frac{n_i^\mu}{n_{ij}} k^- + k_\perp^\mu\,,
 \quad\quad {\rm   \,  denoted \, \, as}  \quad k^\mu = ( k^+, k^-, k_\perp^i ) \,,
 \eea
with $k^+ = n_i \cdot k $, $k^- = n_j \cdot k$ and $k_\perp \cdot n_i = k_\perp \cdot n_j = 0$ and
$| k_\perp |^2 = \frac{2}{n_{ij}} k^+k^- $. Here
 we have introduced the notation $n_{ij} \equiv n_i \cdot n_j$.

\subsection{single real emission phase space parameterization}

With 4 reference vectors, $n_i$, $n_j$, $n_k$ and $n_l$, at hand, for one real emission with momentum $q_1^\mu$ for computing the NLO soft function or the NNLO real-virtual contribution, we can parameterize
all the involved vectors with respect to $n_i$ and $n_j$, which are
\bea\label{eq:para-nlo}
&& n^\mu_k = (n_k^+,n_k^-,0,n_k^\perp \,; 0 ) \,, \nn \\
&& n^\mu_l =  (n_l^+, n_l^-, n_l^\perp s_{\phi_l} , n^\perp_l \, c_{\phi_l} \,; 0)   \,, \nn \\
&& q_1^\mu =  (q_1^+, q_1^-, q_1^\perp s_{\phi_1} c_{\alpha_1} , q_1^\perp \, c_{\phi_1} \,; q_1^\perp s_{\phi_1} s_{\alpha_1} \,, 0 )  \,, 
\eea
where we have used the freedom to choose the azimuthal angle of the vector $n_k$ and the additional $\epsilon$-angles other than $\alpha_1$ to be $0$.
Here $\alpha_1\,, \phi_l\,, \phi_1 \in [0,\pi]$ with 
\bea
c_{\phi_l} = \frac{
n_{ik} n_{jl} +n_{il} n_{jk} - n_{ij} n_{kl} }{ 2 \sqrt{ n_{ik} n_{il} n_{jk}  n_{jl} } } \,.
\eea
Here the non-zero $\phi_l$ makes $n_k$ and $n_l$ span the azimuthal plain and therefore
$\alpha_1$ is necessary for parameterizing the $d$-dimensional momentum $q_1$, since $q_1$ in general lies outside the pain $n_k$ and $n_l$ belong to. 
We note that since the reference vectors $n_i$'s are physical, for arbitrary numbers of the reference vectors, their ``$\perp$"-components will all be lying
 in one azimuthal plain, and thus Eq.~(\ref{eq:para-nlo}) is the most general parameterization for computing the $N$-jettiness function with arbitrary $N$. 
For the processes similar to what we are considering here, the transverse components $n_k^\perp$ and $n_l^\perp$ have to align with each other ($\phi_l = 0$ or $\pi$) due to
the momentum conservation, therefore one can simplify the parameterization by setting $\alpha_1 = 0$ which reduces to the parameterization used
in~\cite{Boughezal:2015eha, Campbell:2017hsw} for computing $1$-jettiness.

Although in general $n_i$ and $n_j$ can be chosen arbitrarily for performing the Sudakov decomposition, in order to correctly isolate the $\epsilon$-poles, it is more useful
to choose $n_i$ so that $n_i \cdot q$ contributes to $\tau$ as set by $\Theta_{N,i}$ while $n_j \cdot q$ appears as one of the singular poles in the denominator of the soft current matrix ${\cal S}$.
With this parameterization, for non-zero $\alpha$, the phase space integration for single emission can be modified to
\bea\label{eq:1-em-ps}
 \int \frac{\mathrm{d}^{d}q_1}{(2\pi)^{d-1}} \, \delta^+(q_1^2) F(q_1) \delta(\tau - n_i \cdot q_1) \theta(n_j \cdot q_1 - n_i \cdot q_1 ) 
&=& \frac{\pi}{4} \, \left( \frac{2}{n_{ij}} \right)^{1-\epsilon} \,
\left( \frac{1}{2\pi} \right)^{3-2\epsilon} \, \tau^{- 1-2\epsilon} \, \nn \\
&&\hspace{-55.ex}\times  \frac{2\pi^{-\epsilon}}{\Gamma(1-\epsilon) } \, \frac{1}{2}  \int_0^1   \, \mathrm{d} x_1 \,
\mathrm{d} x_2 \, 
\mathrm{d} x_3 \,
 x_1^{-1 + \epsilon}  \, 
\frac{-\epsilon}{x_3^{1+\epsilon}} \,
\, 
\left[ 
2^2 s_{\phi_1}^2 (1-x_3) 
\right]^{-\epsilon} \, 
\left[ \tau^2 x_1^{-1}  \right] \, 
\Big[
F(x_3) + F(1-x_3) \Big]\,, 
\eea
where we have made the variable changes
\bea\label{eq:1-em-subs}
q_1^- = \frac{\tau}{x_1} \,, \quad
c_{\phi_1} = \pi \, x_2\,, \quad 
c_{\alpha_1} = 1 - 2x_3 \,.
\eea
It can also be shown that for $\alpha_1=0$, one can further modify the phase space integration to make $\phi_1 \in [0,2\pi]$ when taking into account the properties of the soft current and the $N$-jettiness observable, as has been done in~\cite{Boughezal:2015eha}.  

Eq.~(\ref{eq:1-em-ps}) is our starting point for calculating the single emission contributions to the soft function numerically and 
we list in the Appendix~\ref{app:single-emission} all the final forms of the phase space integrations suitable for direct numerical evaluation of the one emission contributions.

\subsection{double real emission phase space parameterization}

Now we sketch the parameterization for the double real emission with momenta $q_1$ and $q_2$ at NNLO. We want to add on top of Eq.~(\ref{eq:para-nlo}) the momentum $q_2$ parameterization, meanwhile require that the dot product $q_1^\perp \cdot q_2^\perp$ only relies on the 
azimuthal angle between these two vectors, so as to use the non-linear transformation~\cite{Boughezal:2015eha} in a later stage for the purpose of pole isolation. To find the particular parameterization of $q_2$, we first rotate Eq.~(\ref{eq:para-nlo}) around the $n_{k^\perp}$ axis by $ \alpha_1$ to eliminate the $\alpha_1$ dependence in $q_1^\mu$, then we rotate
in the azimuthal plain by $ \phi_1$ to align the $x$-axis with $q_1^\perp$ to further remove the dependence of $\phi_1$ in the parameterization of  $q_1^\mu$. These rotational operations lead to the momentum representation 
\bea
&& q_1^\mu = (q_1^+\,, q_1^-\,, 0 \,, q_1^\perp; 0) \,, \nn \\
&& n_k^\mu = (n_k^+\,, n_k^-\,, - n_k^\perp s_{\phi_1} \,, n_k^\perp c_{\phi_1} ; 0)  \,, \nn \\
&& n_l^\mu = (n_l^+\,, n_l^-\,, n_l^\perp s^{\alpha_1}_{ \phi_l - \phi_1 } \,, n_l^\perp \, c_{\phi_l  - \phi_1}^{\alpha_1} ; - n_l^\perp s_{\phi_l}\, s_{\alpha_1} ,0 ) \,, 
\eea
where we introduced the notation 
$ c^{\alpha_1}_{\phi_l -  \phi_1} \equiv  c_{\phi_l}  c_{\phi_1} + s_ {\phi_l}  s_{\phi_1} c_{\alpha_1} $. 
It is now straightforward to see that to satisfy our requirements, 
$q_2^\mu$ can be parameterized (in this frame after rotation) as 
\bea
q_2^\mu =
 (q_2^+\,, q_2^-\,, q_2^\perp s_{\phi_2}  c_{\alpha_2}\,, q_2^\perp c_{\phi_2}; 
 q_2^\perp s_{\phi_2}  s_{\alpha_2} c_{\beta } \,, q_2^\perp s_{\phi_2}  s_{\alpha_2} s_{\beta }, 0 ) \,, 
\eea
so that $q^\perp_1 \cdot q^\perp_2 = - |q^\perp_1 ||q_2^\perp |\, c_{\phi_2}$. The $\beta$ angle here is necessary to guarantee
that the $\epsilon$-components of $q_1$ and $q_2$ are not always aligned with each other. Due to the Lorenz invariance 
of the individual emission phase space measure, the rotational operations used to derive the parameterization of $q_2$ do not introduce any complexity to the Jacobian of the phase
space integration. We also note that the parameterization derived here is the most general for computing the double real contributions to any $N$-jettiness soft functions.
As for the processes we are interested in in this work, using the same momentum conservation for the single real emission case, one can set $\alpha_1$ and $\beta$ to $0$ to simplify the phase space integration.

Within this parameterization, the dot product $q_1 \cdot q_2$ takes the form
\bea
q_1 \cdot q_2 = \frac{1}{n_{ij}} \left[ q_1^+ q_2^- + q_1^- q_2^+ - 2 \sqrt{q_1^+ q_2^-  q_1^- q_2^+} c_{\phi_2} \right] \,, 
\eea
which generates a linear singularity in the double real soft current if $q_1$ be collinear to $q_2$ is possible, which prevents us from extracting the $\epsilon$-poles using the Laurent expansion in Eq.~(\ref{eq:Laurent}). 
To extract the poles, the angle $c_{\phi_2}$ is further parameterized as
$c_{\phi_2} = 1-2\eta$, 
with the non-linear parameterization 
\bea
\eta  = \frac{ \left(
\sqrt{q_1^+ q_2^- }  - \sqrt{q_1^- q_2^+ }  \right)^2 (1-x') }
{ \left(
\sqrt{q_1^+ q_2^- }  - \sqrt{q_1^- q_2^+ }  \right)^2
 + 4 \sqrt{q_1^+ q_2^-  q_1^- q_2^+} x' } \,, \quad \quad
 x' =  \sin^2\left( \frac{\pi}{2} x \right) \,,
\eea
if the soft momentum $q_1$ and $q_2$ can be collinear to each other, and 
\bea
\eta = x' \,,  \quad \quad
 x' =  \sin^2\left( \frac{\pi}{2} x \right) \,,
\eea
if $q_1$ and $q_2$ can no way be collinear as constrained by the jettiness measurement, for instance when $\tau = n_i \cdot q_1 + n_j \cdot q_2$. 

Last we comment on the choice of $n_i$ and $n_j$ in the Sudakov decomposition.  The same logic follows the single emission case. The reference vectors $n_i$ and $n_j$ are chosen
if $n_i \cdot q_1$ and $n_j \cdot q_2$ contribute to $\tau$, i.e. $\tau = n_i \cdot q_1 + n_j \cdot q_2$. Otherwise, if both $n_i \cdot q_1$ and $n_i \cdot q_2$ contribute to the jettiness observable, then $n_i$ will be naturally picked as one reference axis for the decomposition, and for convenience, $n_j$ is chosen in the way that the double real soft current is singular as $n_j \cdot q_1 \to 0$ and/or  $n_j \cdot q_2 \to 0$. Additional variable changes are needed to map all variables onto the regime $[0,1]$ suitable for doing the Laurent expansion. The mapping is standard in the sector decomposition calculations and has been detailed in~\cite{Campbell:2017hsw, Li:2018tsq}. After the mapping, in the cases where one encounters the fractional power of the form 
$s^{-\frac{1}{2} + \epsilon}$ with $s\in [0,1]$, a further variable transform $s = \sin^2\left( \frac{\pi}{2} x \right)$, with $x\in [0,1]$, is then used. 
 
\section{Numerical Results}\label{sec:num}
In this section, we present our main results. We highlight our results using the $2$-jettiness soft function to ${\cal O}(\alpha_s^2)$ in $pp \to jj$. Other processes such as $e^+e^- \to 4j$ can be easily obtained by changing properly the values
of $\lambda_{ij}$ in Eq.~(\ref{eq:RV-triple}), without doing any additional integrations.
The $2$-jettiness soft function
depends on four hard directions, which we denote as $n_1$, $n_2$, $n_3$ and $n_4$. We align $n_1$ and $n_2$
with the incoming beam axes in the $\pm z$-directions, and let $n_3$ and $n_4$ lie along the outgoing jet
directions. For the purpose of presenting, though not at all required in our calculation, we let the incoming partons carry exactly the same energy in the collision and show the $2$-jettiness soft function as a function of $n_{13}$ and we present the results in the Laplacian space $\int_0^\infty \mathrm{d} \tau \, e^{-\tau \, z} \sigma(\tau) $. 

We start from the NLO soft function. At NLO, the renormalized $2$-jettiness soft function can be written as a sum of dipole terms 
\bea
S^{(1)} =  \frac{\alpha_s}{4\pi} \sum_{i \ne j} T_i \cdot T_j 
\left(
4 C^{(1,-1)}_{ij} \tilde{L}^2 \, 
+  
4 C^{(1,0)}_{ij} \tilde{L} \, 
+ \left[
 2 C^{(1,1)}_{ij} + \frac{2}{3} \pi^2 C^{(1,-1)}_{ij} \right]
\right) 
\equiv \sum_{i\ne j}T_i \cdot T_j S_{ij}^{(1)} \,, \quad 
\eea
where $C_{ij}^{(1,i)}$ are the coefficients of the $\epsilon^{i}$-poles returned by our calculation using the method for single emission as sketched before. The coefficients
of the logarithmic terms $\tilde{L}^2$ and $\tilde{L}$ can be predicted by solving the renormalization group equation (RGE) of the soft function and therefore serve as checks of our numerical
approach. 

\begin{figure}[!h]
\begin{center}
\includegraphics[width=0.45\textwidth,angle=0]{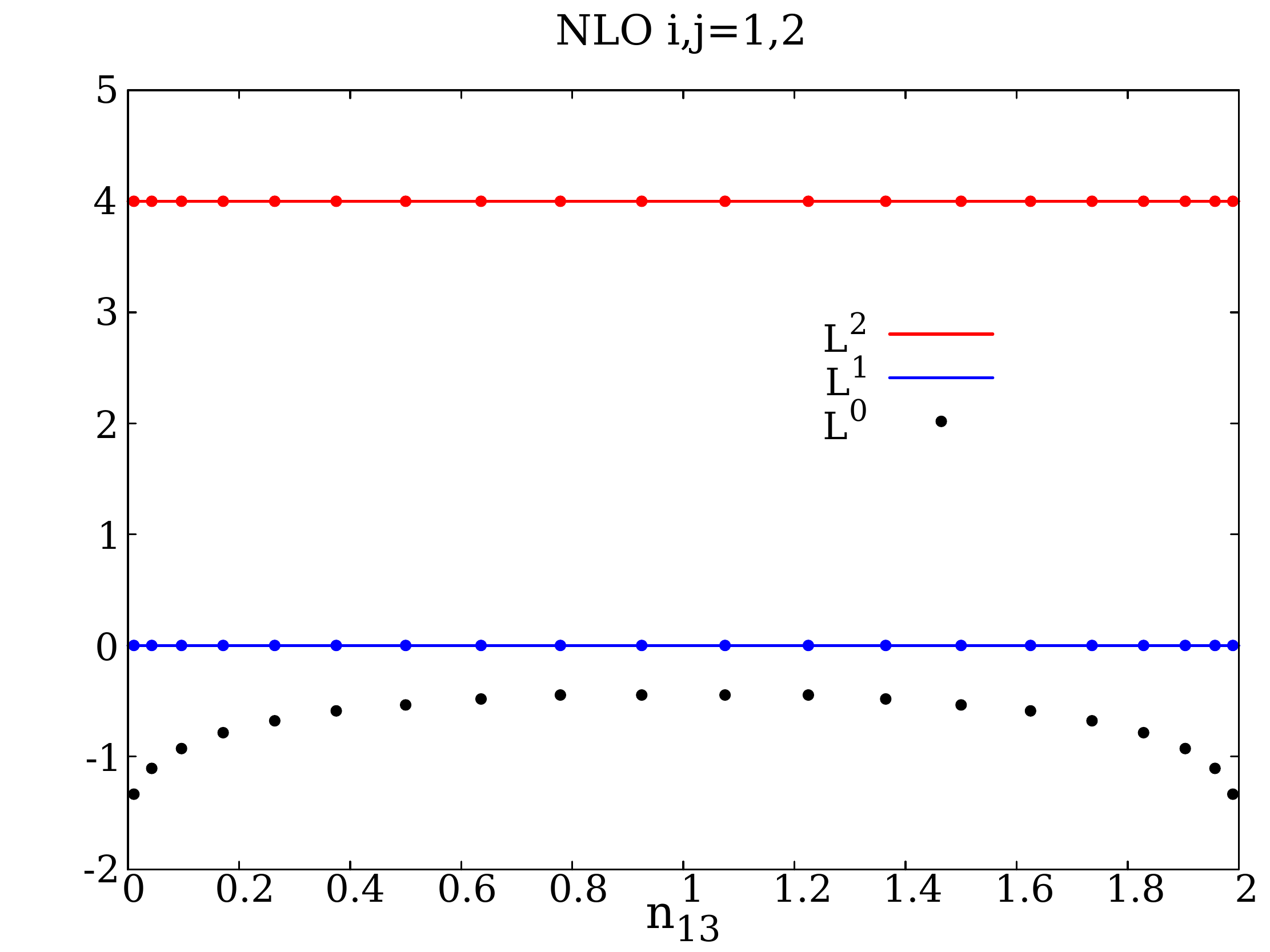} 
\quad \quad
\includegraphics[width=0.45\textwidth,angle=0]{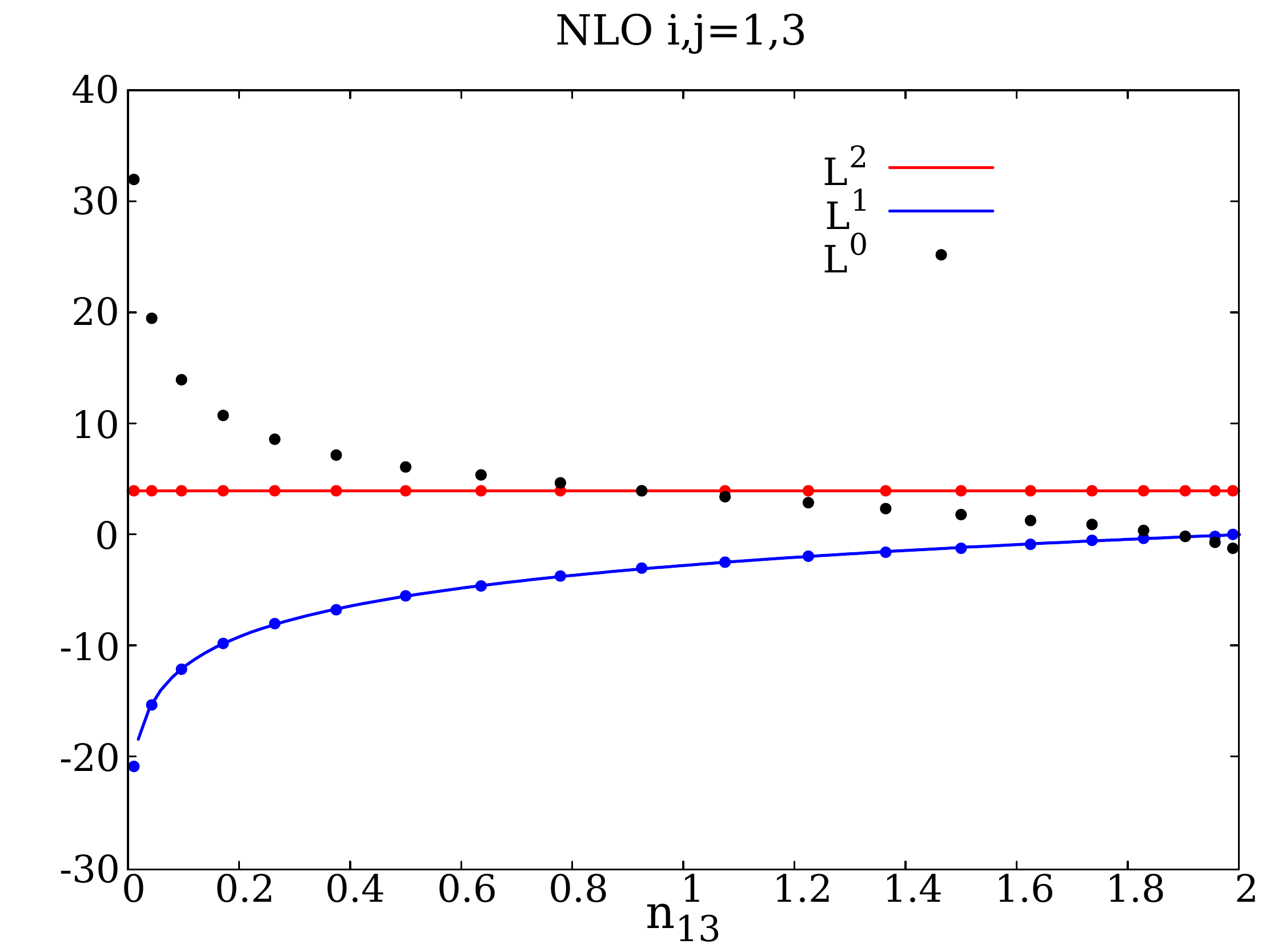}
\end{center}
\vspace{-0.5cm}
\caption{Coefficients of the $\tilde{L}^n$ terms for the ${\cal O}(\alpha_s)$ $2$-jettiness soft function in $pp \to jj$ production, normalized to 
$\frac{\alpha_s}{4\pi} T_i \cdot T_j $. } \label{fig:NLO}
\end{figure}

 In fig.~\ref{fig:NLO}, we show the NLO coefficients of the $\tilde{L}^n$ ($n = 2,1,0$) terms for dipole $S_{12}^{(1)}$ and $S_{13}^{(1)}$ as functions of $n_{13}$. These are the only independent terms
 in the special kinematics we have chosen. 
In both cases, we have normalized the results to $\frac{\alpha_s}{4\pi} $. The red, blue and black dots represent the coefficients for $\tilde{L}^2$, 
$\tilde{L}^1$ and $\tilde{L}^0$ from our numerical set-ups, respectively. The solid lines are obtained by solving the RGE analytically. We see perfect agreements between
the analytic results and our numerical predictions, which implies the validity of the computational framework at this order.
 
At NNLO, other than the abelian terms which can be obtained by exponentiate the NLO results, the soft function receives non-abelian contributions from the $q{\bar q}$-emission, the $gg$-emission, the real-virtual corrections and the renormalization. The real-virtual corrections can be further broke down to the dipole and the triple-pole contributions, and the rest are of the dipole forms. The summation of all the dipole contributions can be organized by the color factors $C_A$ and $N_F T_R$, while the triple-pole real-virtual corrections can be massaged into one single color term $f_{abc} T^a_1 T^b_2 T^c_3$ with the help of the color charge conservation $\sum T_i = 0$. The final results of the NNLO soft function can therefore be written as
\bea
S^{(2)} &=&\frac{1}{4} \sum_{i\ne j, k\ne l} \{T_i \cdot T_j, T_k \cdot T_l \} \, S^{(1)}_{ij} S^{(1)}_{kl} \nn \\
&& + \left( \frac{\alpha_s}{4\pi} \right)^2 \, \left(
\sum_{i\ne j}T_i\cdot T_j \left[ C_A  S_{ij}^{(2),C_A} + N_F T_R S_{ij}^{(2),N_F} \right]  
+ f_{abc} T_1^a T_2^b T_3^c \, S_{123}^{(2)} \right) \,,
\eea
where all the NNLO dipoles $S_{ij}^{(2)}$ and triple-pole $S_{123}^{(2)}$ in the non-abelian terms can be written
as a series in terms of the logarithms $\tilde{L}$. 
And once again, all the coefficients of those logarithmic terms $\tilde{L}^n$ with $n = 3,2,1$ can be predicted by RGE with the 
knowledge of the full NLO results obtained before. 

\begin{figure}[!h]
\begin{center}
\includegraphics[width=0.45\textwidth,angle=0]{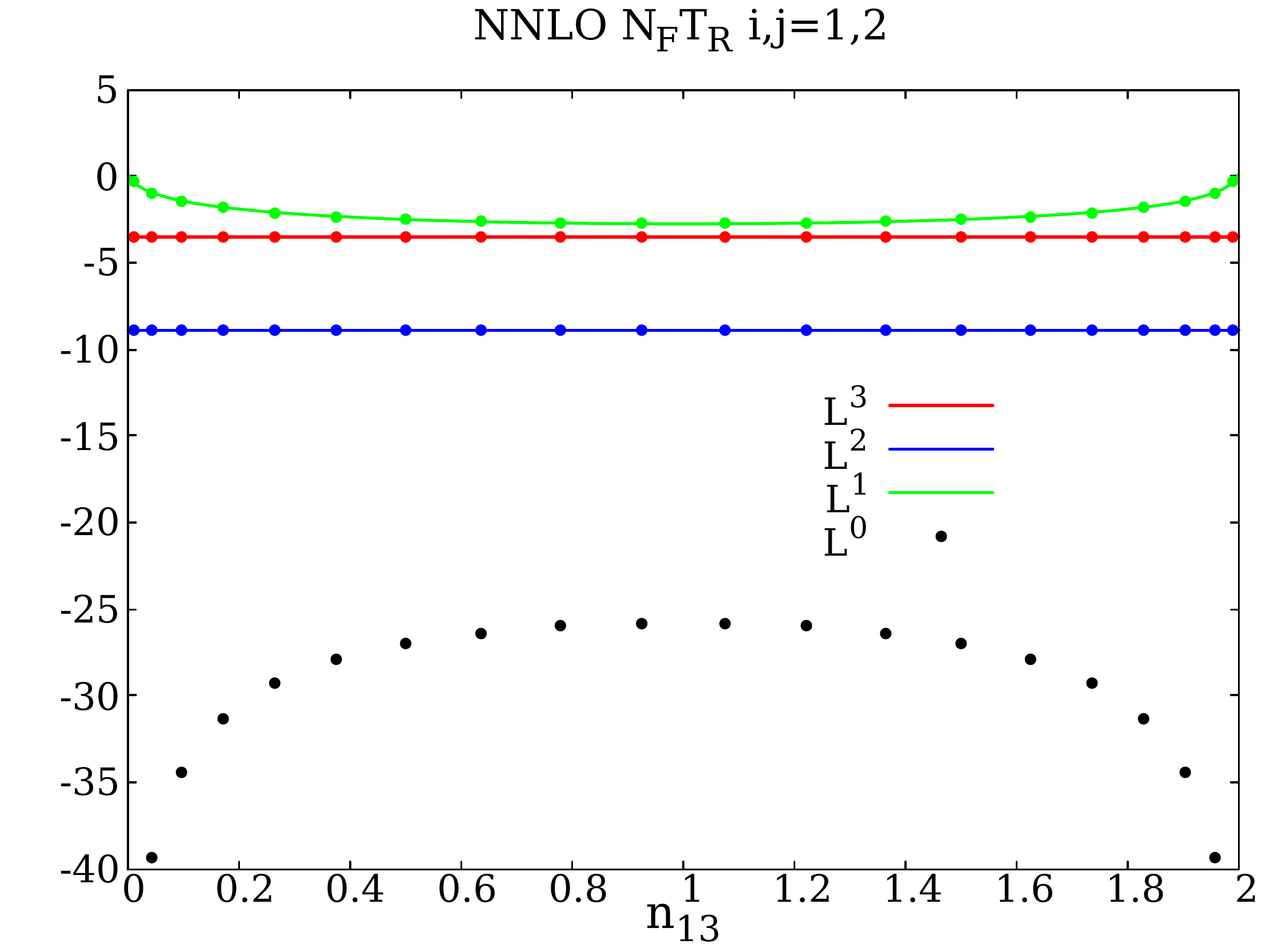} 
\quad \quad
\includegraphics[width=0.45\textwidth,angle=0]{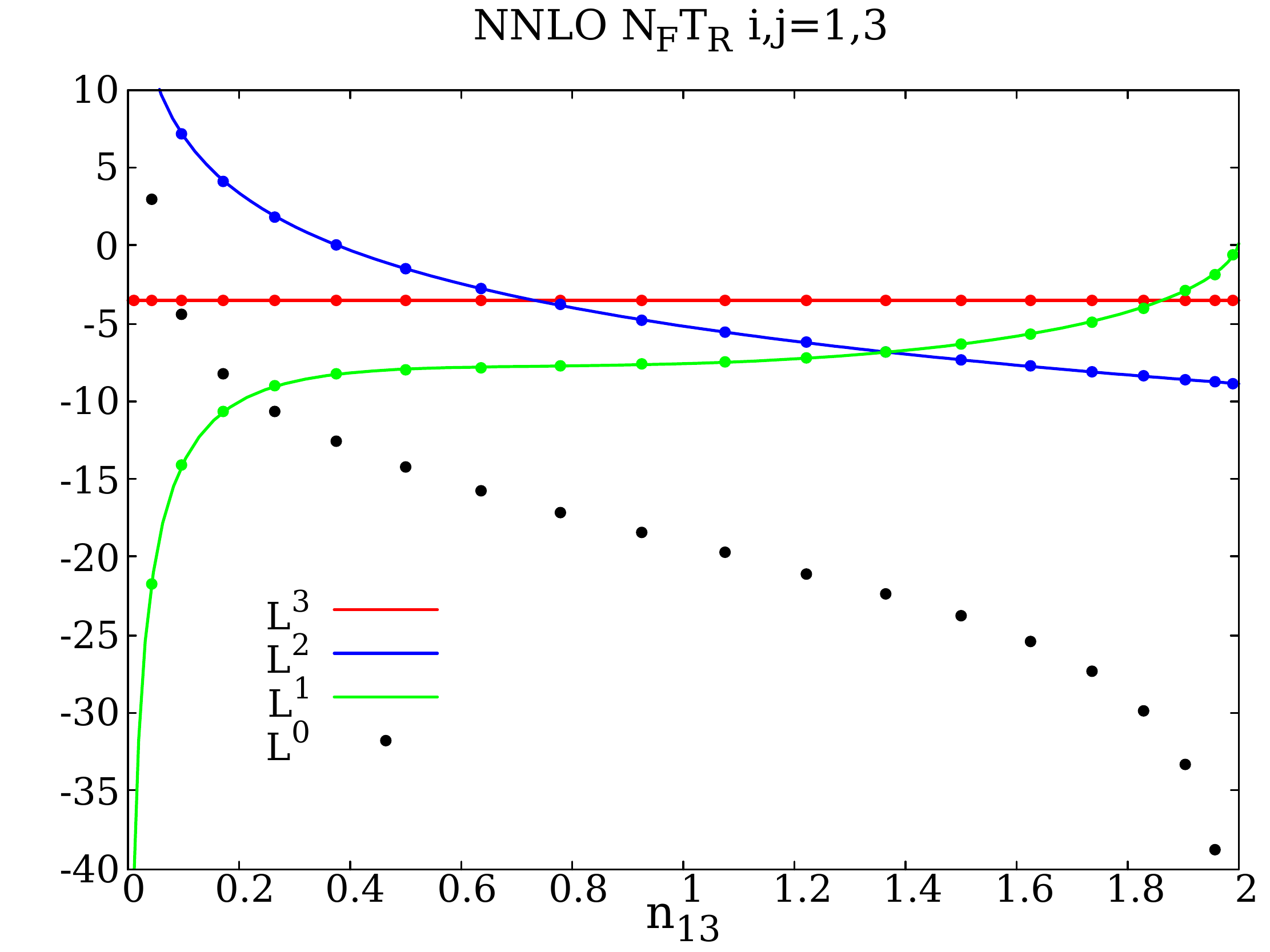}
\end{center}
\vspace{-0.5cm}
\caption{Coefficients of the $\tilde{L}^n$ with $n = 3,2,1,0$ in $S_{12}^{(2),N_F}$ (left panel) and $S_{13}^{(2),N_F}$ (right panel). } \label{fig:NNLO-NF}. 
\end{figure}

The coefficients of the $N_F T_R$ terms for the dipoles $S_{12}^{(2)}$ and $S_{13}^{(2)}$ are shown in fig.~\ref{fig:NNLO-NF}, which include
the contributions from the $q{\bar q}$ emission and the $N_F$ term in the renormalization.
Perfect agreements are observed between the numerical calculations of the $\tilde{L}^n$ coefficients for $n = 3,2,1$ as shown in colored dots and the RGE predictions in solid lines, which validates not only the NNLO computations of the $N_F$ contributions but also the correctness of the 
non-logarithmic term in the NLO calculations. The NNLO non-logarithmic prediction (in black dots) displayed here is not presented in the original NNLO $N$-jettiness soft paper~\cite{Boughezal:2015eha} and is one of our main results of this work. We note that the NNLO results here and below are also presented in a conference proceeding~\cite{Bell:2018mkk} in a different way with a different approach.  
 \begin{figure}[!h]
\begin{center}
\includegraphics[width=0.45\textwidth,angle=0]{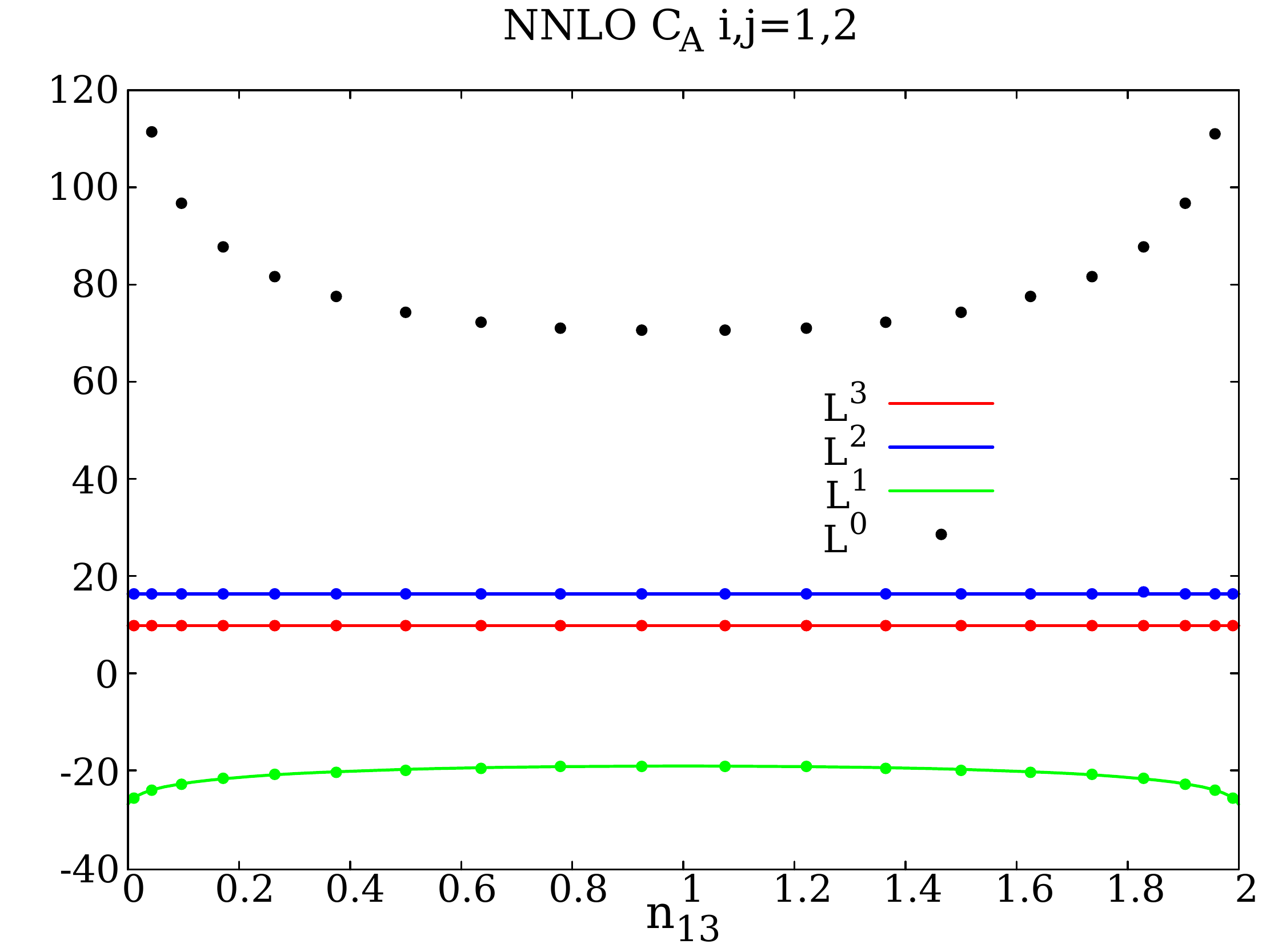} 
\quad \quad
\includegraphics[width=0.45\textwidth,angle=0]{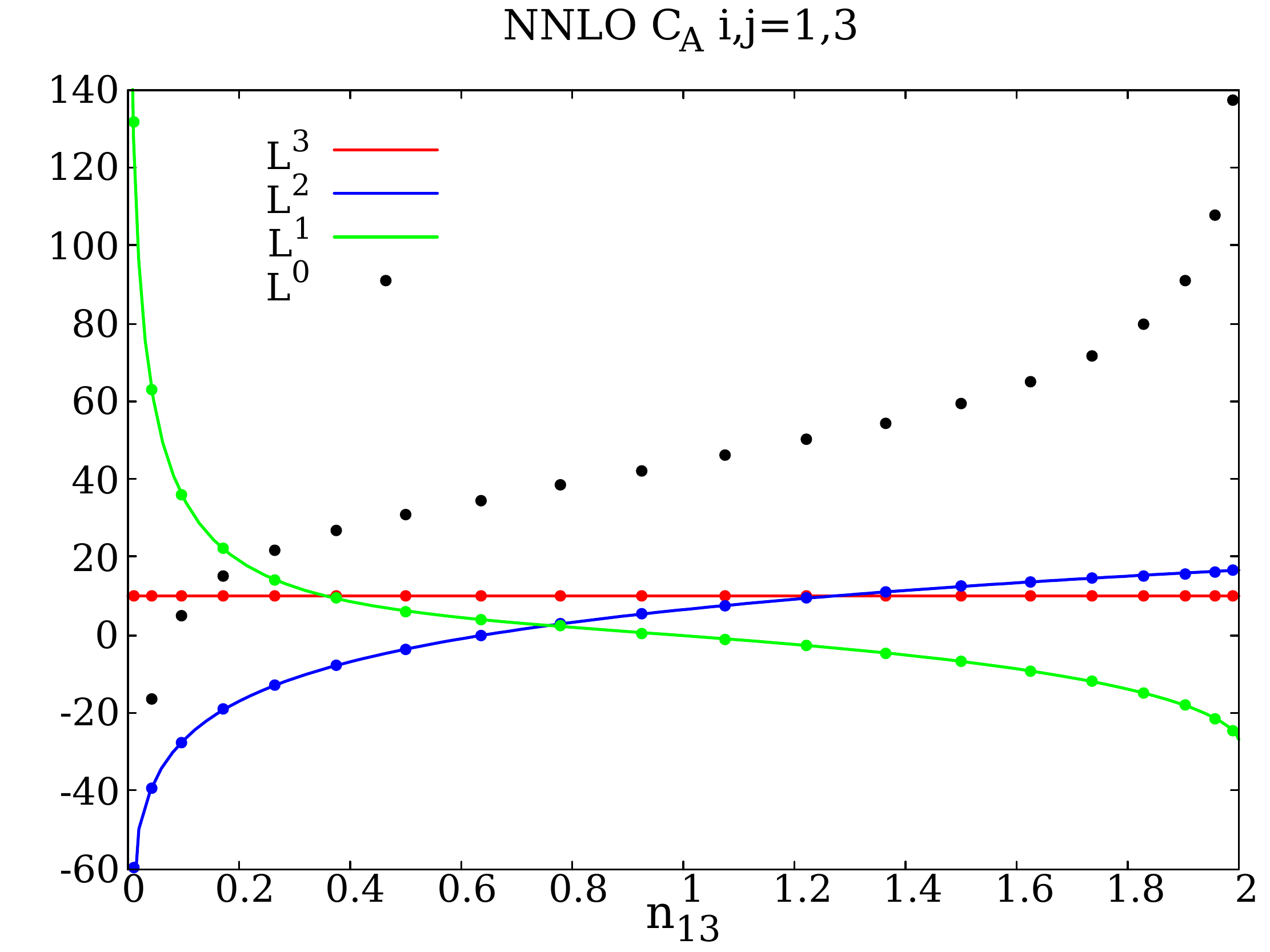}
\end{center}
\vspace{-0.5cm}
\caption{Coefficients of the $\tilde{L}^n$ with $n = 3,2,1,0$ in $S_{12}^{(2),C_A}$ (left panel) and $S_{13}^{(2),C_A}$ (right panel). } \label{fig:NNLO-CA}. 
\end{figure}

Same situation happens to the $C_A$ term, as displayed in fig.~\ref{fig:NNLO-CA}, in which we also compared $\tilde{L}^n$ coefficients from the numerical calculation against the results from the RGE to find no difference. The non-logarithmic coefficients are represented by the black dots.
 \begin{figure}[!h]
\begin{center}
\includegraphics[width=0.45\textwidth,angle=0]{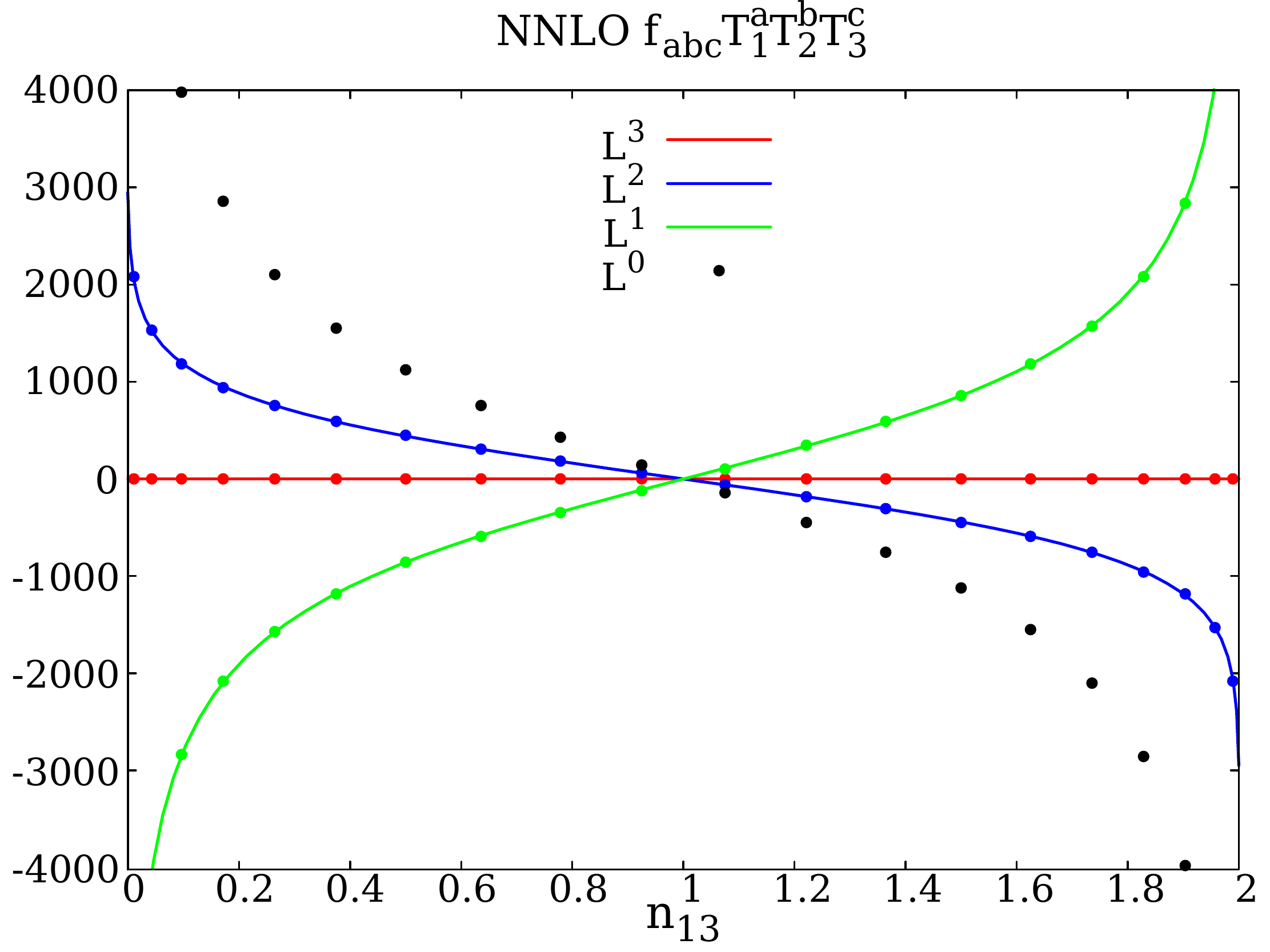}  
\quad \quad
\includegraphics[width=0.45\textwidth,angle=0]{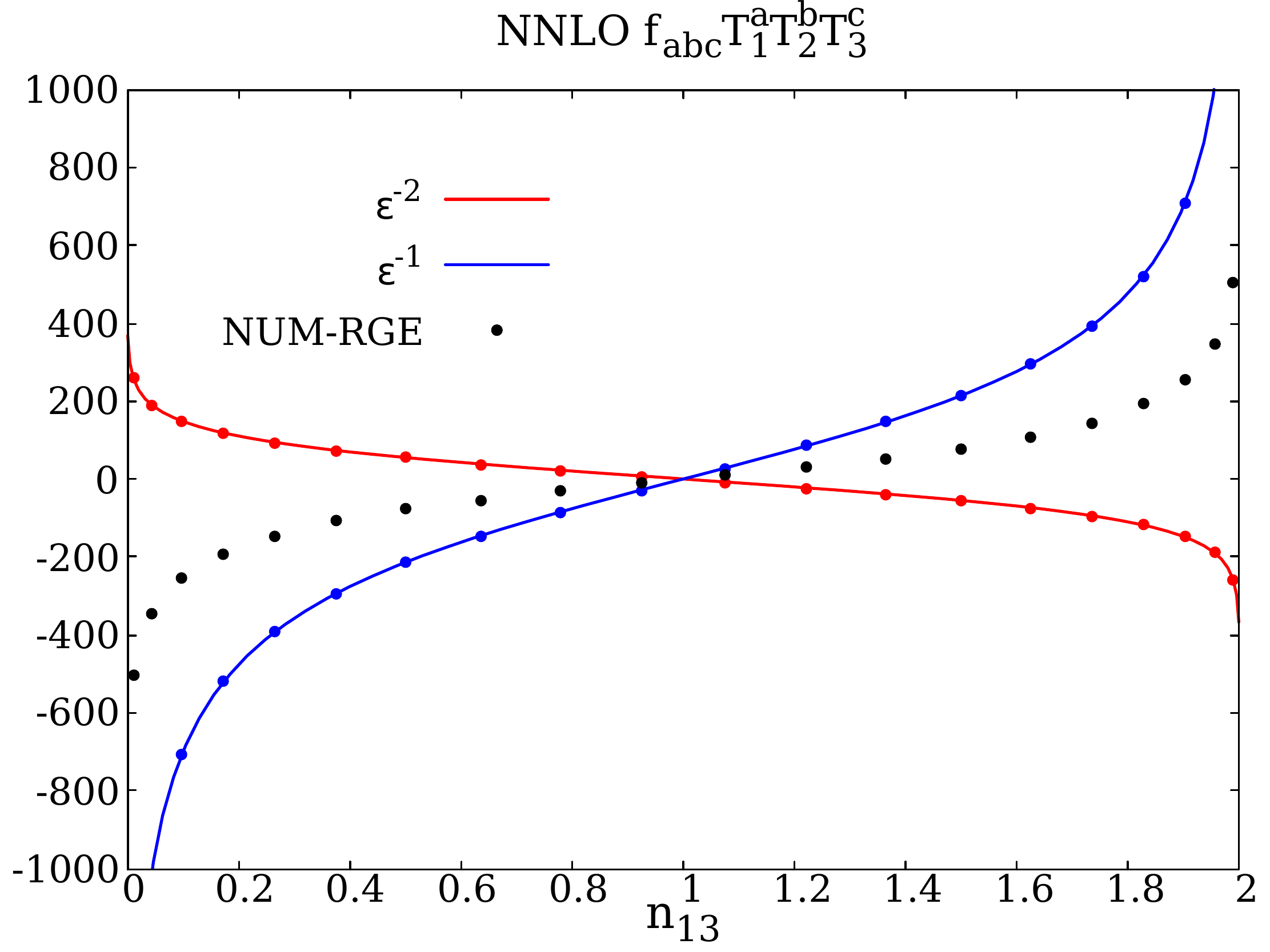} 
\end{center}
\vspace{-0.5cm}
\caption{Coefficients of the $\tilde{L}^n$ with $n = 3,2,1,0$ in $S_{123}^{(2)}$ (left panel) and the $\epsilon$-pole terms (right panel). } \label{fig:NNLO-TTT}. 
\end{figure}

Last we give the prediction for the triple-pole term $S_{123}^{(2)}$, which arises firstly for four hard reference directions. Just like the dipole contributions, the logarithms in the triple-pole are still predictable from the RGE. The analytic predictions are once again found to agree with our numerical results, as clearly seen in the left panel of fig.~\ref{fig:NNLO-TTT}. Other than that, we also compare the $\epsilon$-pole terms without the $\tilde{L}$'s from our numerical calculations with the ones predicted by the RGEs and find satisfactory agreements. The result for the non-logarithmic term (left panel in fig.~\ref{fig:NNLO-TTT}) as well as the term entirely unpredictable from RGE (right panel, normalized further to $-2\pi$) are indicated in black dots, which finalizes all the NNLO contributions to the $2$-jettiness for the $pp \to jj$ process.


\section{Conclusions}\label{sec:conclude}
In this work, we present the calculation of the $N$-jettiness soft function with four hard reference directions using the computational framework proposed in~\cite{Boughezal:2015eha} using the approach of sector decomposition. We generalize the parameterizations used for the $1$-jettiness calculation to the arbitrary $N$-jettiness case. We specifically calculated 
the NNLO $N$-jettiness soft function with four external hard legs, where the last new piece at this order, the triple-pole configuration, first arises. We managed to isolate all the singularities from the soft function integrands and reduce the computation to a set of numerical integrals which are ready to evaluate. We check the numerically computed logarithmic terms in the soft function against the predictions from the RGEs and found perfect agreements. The non-logarithmic terms which can not be obtained through the known RGEs are calculated directly in this work.  We expect the result obtained in this manuscript to find its quick applications in both the fixed NNLO calculations and achieving the parton shower matching~\cite{Alioli:2013hqa} for the relevant processes. 

\section{Acknowledgments}
This work is supported by the National Natural Science Foundation of China under Grant No. 11775023
and the Fundamental Research Funds for the Central Universities. S.J. is supported by the BNU JingShi Undergraduate Fellowship.

\section*{Appendix}
\subsection{soft current}\label{app:soft-current}
For completeness, we list all the soft current matrix elements here, which were first derived in the seminal works~\cite{Catani:1999ss} and~\cite{Catani:2000pi} by Catani and Grazzini. 

The NLO soft current with one soft emission $q$ is given by
\bea
{\cal S}^{(1)} = - (4\pi \alpha_s \mu^{2\epsilon})  \times 2 \times \sum_{i\ne j}  
T_i \cdot T_j  \, {\cal S}^{(1)}_{ij}(q) \,,
\eea
with the NLO dipole given by
\bea
{\cal S}^{(1)}_{ij}(q) = \frac{n_{ij}}{2\, n_i \cdot q \, n_j \cdot q} \,. 
\eea

The NNLO real-virtual correction can be written as the sum of the dipole ${\cal S}^{(2)}_{RV,dipole}$
and triple-pole ${\cal S}^{(2)}_{RV,tri.}$ contributions, with the 
dipole contribution be
\bea
{\cal S}^{(2)}_{RV,dipole} = 
\frac{ (4\pi\alpha_s\, \mu^{2\epsilon} )^2}{4\pi^2} \, 
\times 
 \frac{(4\pi )^\epsilon}{\epsilon^2} \,
\frac{ \Gamma^4(1-\epsilon) \, \Gamma^3(1+\epsilon) }
{\Gamma^2(1-2\epsilon)\, \Gamma(1+2\epsilon) } \,  \times \, 
C_A\, \,
\sum_{i \ne j}
T_i \cdot T_j \, 
\left[ {\cal S}^{(1)}_{ij}(q)  \right]^{1+\epsilon} \,,
\eea
and the triple-pole term 
\bea\label{eq:RV-triple}
{\cal S}^{(2)}_{RV,tri.} &=&  \frac{ ( 4\pi \alpha_s \mu^{2\epsilon} )^2 }{4\pi^2}
\times
 \frac{(4\pi)^\epsilon}{\epsilon^2 }\,
\frac{\Gamma^3(1-\epsilon) \Gamma^2(1+\epsilon)}{\Gamma(1-2\epsilon)}\,
2 \sin(\pi \epsilon) \nn \\
&& \times  \sum_{i\ne j\ne k} \, f_{abc} T_k^a \, T_i^b \, T_j^c\, \,
\left(
\lambda_{ij} - \lambda_{iq} - \lambda_{jq}
\right) \, \,
{\cal S}_{ki}(q) \left[ {\cal S}_{ij}(q) \right]^\epsilon \,
\,, 
\eea
where $\lambda_{ij} = 1$ if $i$ and $j$ are both incoming or outgoing,
otherwise $\lambda_{ij} = 0$.

The NNLO real-real emission is given by 
\bea
{\cal S}_{q{\bar q}}^{(2)} = 
-\frac{1}{2} (4\pi \alpha_s \mu^{2\epsilon} )^2 \, T_R N_F \sum_{i\ne j} T_i \cdot T_j \, {\cal J}_{ij} \,, 
\eea
from the $q{\bar q}$ emission. To get this form we have used the color charge conservation $\sum_i T_i = 0$. 
Here 
\bea
{\cal J}_{ij} = {\cal I}_{ii} + {\cal I}_{jj} - 2 {\cal I}_{ij} \,, 
\eea
and 
\bea
{\cal I}_{ij} = - \frac{2 n_{ij} \, q_1  \cdot q_2 + \, n_i\cdot(q_1-q_2) \, n_j\cdot(q_1-q_2) }
{2 (q_1\cdot q_2)^2  \, \, n_i \cdot (q_1+q_2) n_j\cdot (q_1+q_2) } \,, 
\eea
where $q_1$ and $q_2$ are the momentum carried by the real emissions. 
We denote the term proportional to $2 \, n_{ij} \,  q_1\cdot q_2$ as ${\cal I}_{ij}^{\rm II}$ and the rest 
${\cal I}_{ij}^{\rm I}$. 

And the $gg$ emission gives the non-abelian contribution
\bea
{\cal S}_{gg,non-abe.}^{(2)} = - \frac{1}{2} \, 
\left( 4\pi \alpha_s \mu^{2\epsilon} \right)^2 \, C_A \, \sum_{i\ne j } \, T_i\cdot T_j\, {\cal T}_{ij} \,,
\eea
with 
\bea
{\cal T}_{ij} = 2 {\cal S}_{ij} - {\cal S}_{ii} - {\cal S}_{jj}\,,
\eea
where 
\bea
{\cal S}_{ij} = {\cal S}_{ij}^{\rm s.o.} - \frac{1}{2} \,
\frac{n_i \cdot q_1 \, n_j \cdot q_2 + n_i \cdot q_2 \, n_j \cdot q_1}{n_i\cdot(q_1+q_2) \, n_j\cdot(q_1+q_2) } \,
{\cal S}_{ij}^{\rm s.o.} + 
(1-\epsilon) \,  {\cal I}_{ij}^{\rm I}  + 2 \, {\cal I}_{ij}^{\rm II} \,,
\eea
and the current in the strongly ordering limit  
\bea
{\cal S}_{ij}^{\rm s.o.} &=&
\frac{n_{ij}}{q_1\cdot q_2} \, \left(
\frac{1}{n_i \cdot q_1 \, n_j \cdot q_2} + 
\frac{1}{n_j \cdot q_1 \, n_i \cdot q_2}
\right)
- \frac{n_{ij}^2}{ n_i \cdot q_1 \, n_i \cdot q_2 \, n_j \cdot q_1 \, n_j \cdot q_2 } \,. 
\eea
The abelian piece for the $gg$ double real emission is 
\bea
{\cal S}_{gg,abelian}^{(2)}
= 4 \times (4\pi\alpha_s \mu^{2\epsilon})^2 \times \frac{1}{2} \sum_{i\ne j,k\ne l} \{T_i \cdot T_j \,,
T_k \cdot T_l
\} {\cal S}_{ij}^{(1)}(q_1) \,  {\cal S}_{kl}^{(1)}(q_2) \,.  
\eea

\subsection{numerical integrations for single emission}\label{app:single-emission}
Here we list all final numerical integrations for evaluate single emission contributions. We note again
that though we only show the $N$-jettiness with four hard directions $n_i$, $n_j$, $n_k$ and $n_l$, 
 the parameterization here presented here is general enough for any $N$-jettiness soft functions. To extend
 to the case with more reference vectors, one simply insert more $\theta$ functions originated from the 
 $N$-jettiness measurements. We use the notation $q || n_i$ to denote the separation of $q$ and $n_i$
 is the smallest and thus $\tau = n_i \cdot q$.

\subsubsection{final forms for numerical integration at NLO}
At NLO, the contribution can be written as a sum of dipoles. We assume that the soft momentum $q$
is emitted from dipole $ij$. For the integrations shown below, we have extracted out an overall factor 
$ - 16\pi^2\left(\frac{\alpha_s}{2\pi}\right)\left(\frac{e^{\gamma_E}}{4\pi}\right)^\epsilon T_i \cdot T_j$.

\begin{itemize}
\item \textbf{case 1}: $q_1\parallel n_i$.

For $q_1\parallel n_i$, the final form suitable for numerical integration is
\begin{eqnarray}
I^{i,{(1)}}_{ij} &= &\frac{\pi}{4}   \left( \frac{n_{ij}}{2} \right)^{\epsilon} 
\left( \frac{1}{2\pi} \right)^{3-2\epsilon}  \tau^{- 1-2\epsilon} 
 \frac{2\pi^{-\epsilon}}{\Gamma(1-\epsilon) }  \frac{1}{2}  \int_0^1    \mathrm{d} x_1 
\mathrm{d} x_2 
\mathrm{d} x_3 \,
 x_1^{-1 + \epsilon}  
\frac{-\epsilon}{x_3^{1+\epsilon}}  
\left[ 
2^2 s_{\phi_1}^2 (1-x_3) 
\right]^{-\epsilon} \nonumber\\
&& \times \Big[
F^{ij}_i(x_3) + F^{ij}_i(1-x_3) \Big] \label{eq:1-em-num-1i} \,, 
\end{eqnarray}
where we have defined
\begin{equation}
F^{ij}_i(x_3)=\theta\big(A_{ij,k}(x_1,c_{\phi_1})-x_1\big)
\theta\big(A_{ij,l}(x_1,c^{\alpha_1}_{\phi_l -  \phi_1})-x_1\big) \,, 
\end{equation}
and
\begin{equation}
	A_{ij,k}(x,c_\phi)=\frac{n_{ik}}{n_{ij}}+\frac{n_{jk}}{n_{ij}}x-\sqrt{\frac{2x}{n_{ij}}}n_k^\perp c_\phi \,.
\end{equation}

As for the case in which $q_1\parallel n_j$, the result can be obtained by switching $i$ and $j$ in Eq.~(\ref{eq:1-em-num-1i}).

\item \textbf{case 2}: $q_1\parallel n_k$. 

While for $q_1\parallel n_k$, we choose $n_k$ and $n_i$ to be the reference vectors, to have $q_1^+=n_k\cdot q_1,q_1^-=n_i\cdot q_1$. Following the same variable changes shown in Eq.~(\ref{eq:1-em-subs}), we get the final form for the numerical evaluation 
\begin{eqnarray}
I^{k,{(1)}}_{ij}&=&\frac{\pi}{4}  \frac{n_{ij}}{n_{ik}}  \left( \frac{n_{ik}}{2} \right)^{\epsilon}
\left( \frac{1}{2\pi} \right)^{3-2\epsilon}  \tau^{- 1-2\epsilon} 
 \frac{2\pi^{-\epsilon}}{\Gamma(1-\epsilon) }  \frac{1}{2}  \int_0^1    \mathrm{d} x_1 
\mathrm{d} x_2 
\mathrm{d} x_3 \,
 x_1^{\epsilon}  
\frac{-\epsilon}{x_3^{1+\epsilon}}  
\frac{
\left[ 
2^2 s_{\phi_1}^2 (1-x_3) 
\right]^{-\epsilon}
}{A_{ki,j}(x_1,c_{\phi_1})} \nonumber\\
&& \times \Big[
F^{ij}_k(x_3) + F^{ij}_k(1-x_3) \Big]  \,, 
\end{eqnarray}
where
\begin{equation}
F^{ij}_k(x_3) =	\theta\big(A_{ki,j}(x_1,c_{\phi_1})-x_1\big)
\theta\big(A_{ki,l}(x_1,c^{\alpha_1}_{\phi_l -  \phi_1})-x_1\big) \,.
\end{equation}
Again, we can switch $k$ and $l$ to obtain the $q_1\parallel n_l$ contribution.

\end{itemize}

\subsubsection{final forms for $2$-parton correlated real-virtual contribution}
For the $2$-parton correlated real-virtual contribution, we have for our numerical evaluation 

\begin{itemize}

\item \textbf{case 1}: $q_1\parallel n_i$. 
\begin{eqnarray}
I^{i,{(2)}}_{ij,{\rm RV}} &=& \frac{\pi}{4}   \left( \frac{n_{ij}}{2} \right)^{2\epsilon} 
\left( \frac{1}{2\pi} \right)^{3-2\epsilon}  \tau^{- 1-4\epsilon} 
 \frac{2\pi^{-\epsilon}}{\Gamma(1-\epsilon) }  \frac{1}{2}  \int_0^1    \mathrm{d} x_1 
\mathrm{d} x_2 
\mathrm{d} x_3 \,
 x_1^{-1 + 2\epsilon}  
\frac{-\epsilon}{x_3^{1+\epsilon}}  
\left[ 
2^2 s_{\phi_1}^2 (1-x_3) 
\right]^{-\epsilon} \nonumber\\
&& \times \Big[
F^{ij}_i(x_3) + F^{ij}_i(1-x_3) \Big]  \,,  
\end{eqnarray}

\item \textbf{case 2:} $q_1\parallel n_k$. 
\begin{eqnarray}
I^{k,{(2)}}_{ij,{\rm RV}} &=& \frac{\pi}{4}  \frac{n_{ij}}{n_{ik}}  
\left( \frac{n_{ij}n_{ik}}{4} \right)^{\epsilon}
\left( \frac{1}{2\pi} \right)^{3-2\epsilon}  \tau^{- 1-4\epsilon} 
 \frac{2\pi^{-\epsilon}}{\Gamma(1-\epsilon) }  \frac{1}{2}  \int_0^1    \mathrm{d} x_1 
\mathrm{d} x_2 
\mathrm{d} x_3 \,
 x_1^{3\epsilon}  
\frac{-\epsilon}{x_3^{1+\epsilon}}  
\frac{
\left[ 
2^2 s_{\phi_1}^2 (1-x_3) 
\right]^{-\epsilon}
}{\Big[A_{ki,j}(x_1,c_{\phi_1})\Big]^{1+\epsilon}} \nonumber\\
&& \times \Big[
F^{ij}_k(x_3) + F^{ij}_k(1-x_3) \Big]  \,.
\end{eqnarray}

\end{itemize}
Here we have normalized to 
\bea
16\pi^2\left(\frac{\alpha_s}{2\pi}\right)^2 \left(\frac{e^{\gamma_E}}{4\pi}\right)^{2\epsilon}\frac{(4\pi)^\epsilon}{\epsilon^2}\frac{\Gamma^4(1-\epsilon)\Gamma^3(1+\epsilon)}{\Gamma^2(1-2\epsilon)\Gamma(1+2\epsilon)} C_A T_i \cdot T_j \,. 
\eea
The $q || n_j$ and $q || n_l$ cases can be easily obtained by switching properly the indices. 

\subsubsection{final forms for 3-parton correlated real-virtual contribution}

Now we turn to the triple-pole case in the real-virtual correction, in which $3$-parton correlated emission contributes. This 
configuration first arises in this case with four external legs. We assume the soft
emission is from triple-pole $ijk$. And we normalize our results to 
\bea
32\pi^2\left(\frac{\alpha_s}{2\pi}\right)^2\left(\frac{e^{\gamma_E}}{4\pi}\right)^{2\epsilon}\frac{(4\pi)^\epsilon}{\epsilon^2}\frac{\Gamma^3(1-\epsilon)\Gamma^2(1+\epsilon)}{\Gamma(1-2\epsilon)}\sin(\pi\epsilon)  f_{abc} T^a_k T^b_i T^c_j \left(
\lambda_{ij} - \lambda_{iq} - \lambda_{jq}
\right) \,. 
\eea

\begin{itemize}

\item \textbf{case 1:} $q_1\parallel n_i$. 
\begin{eqnarray}
I_{ijk}^{i,(2)} & = &\frac{\pi}{4}  \frac{n_{ik}}{n_{ij}}  \left( \frac{n_{ij}}{2} \right)^{2\epsilon}
\left( \frac{1}{2\pi} \right)^{3-2\epsilon}  \tau^{- 1-4\epsilon} 
 \frac{2\pi^{-\epsilon}}{\Gamma(1-\epsilon) }  \frac{1}{2}  \int_0^1    \mathrm{d} x_1 
\mathrm{d} x_2 
\mathrm{d} x_3 \,
 x_1^{-1+2\epsilon}  
\frac{-\epsilon}{x_3^{1+\epsilon}}  
\frac{
\left[ 
2^2 s_{\phi_1}^2 (1-x_3) 
\right]^{-\epsilon}
}{A_{ij,k}(x_1,c_{\phi_1})} \nonumber\\
&& \times \Big[
F^{ij}_i(x_3) + F^{ij}_i(1-x_3) \Big] \,,
\end{eqnarray}
where $F^{ij}_i$ has been defined before. 

\item \textbf{case 2:} $q_1\parallel n_j$

We let $q_1^+=n_j\cdot q_1$, $q_1^-=n_i\cdot q_1$ to find
\begin{eqnarray}
I_{ijk}^{j,(2)} &=&\frac{\pi}{4}  \frac{n_{ik}}{n_{ij}}  \left( \frac{n_{ij}}{2} \right)^{2\epsilon}
\left( \frac{1}{2\pi} \right)^{3-2\epsilon}  \tau^{- 1-4\epsilon} 
 \frac{2\pi^{-\epsilon}}{\Gamma(1-\epsilon) }  \frac{1}{2}  \int_0^1    \mathrm{d} x_1 
\mathrm{d} x_2 
\mathrm{d} x_3 \,
 x_1^{2\epsilon}  
\frac{-\epsilon}{x_3^{1+\epsilon}}  
\frac{
\left[ 
2^2 s_{\phi_1}^2 (1-x_3) 
\right]^{-\epsilon}
}{A_{ji,k}(x_1,c_{\phi_1})} \nonumber\\
&& \times \Big[
F^{ij}_j(x_3) + F^{ij}_j(1-x_3) \Big]  \,. 
\end{eqnarray}

\item \textbf{case 3:} $q_1\parallel n_k$

We let $q_1^+=n_k\cdot q_1$ and $q_1^-=n_l\cdot q_1$ to get
\begin{eqnarray}
I_{ijk}^{k,(2)} &=&\frac{\pi}{4}  \frac{n_{ik}}{n_{kl}}  \left( \frac{n_{ij}n_{kl}}{4} \right)^{\epsilon}
\left( \frac{1}{2\pi} \right)^{3-2\epsilon}  \tau^{- 1-4\epsilon} 
 \frac{2\pi^{-\epsilon}}{\Gamma(1-\epsilon) }  \frac{1}{2}  \int_0^1    \mathrm{d} x_1 
\mathrm{d} x_2 
\mathrm{d} x_3 \,
 x_1^{-1+3\epsilon}  
\frac{-\epsilon}{x_3^{1+\epsilon}}  
\frac{
\left[ 
2^2 s_{\phi_1}^2 (1-x_3) 
\right]^{-\epsilon}
}{\Big[A_{kl,i}(x_1,c_{\phi_1})\Big]^{1+\epsilon}} \nonumber\\
&& \times \Big[
F^{ij}_k(x_3) + F^{ij}_k(1-x_3) \Big]  \,.
\end{eqnarray}

Here we redefine $F^{ij}_k(x_3)$ as
\begin{equation}
	F^{ij}_k(x_3)=\left[\frac{1}{A_{kl,j}(x_1,c^{\alpha_1}_{\phi_j -  \phi_1})}\right]^{\epsilon}
	\theta\big(A_{kl,i}(x_1,c_{\phi_1})-x_1\big)
\theta\big(A_{kl,j}(x_1,c^{\alpha_1}_{\phi_j -  \phi_1})-x_1\big) \,. 
\end{equation}

\item \textbf{case 4:} $q_1\parallel n_l$

We let $q_1^+=n_l\cdot q_1$ and $q_1^-=n_k\cdot q_1$ to have
\begin{eqnarray}
I_{ijk}^{l,(2)} &=&\frac{\pi}{4}  \frac{n_{ik}}{n_{kl}}  \left( \frac{n_{ij}n_{kl}}{4} \right)^{\epsilon}
\left( \frac{1}{2\pi} \right)^{3-2\epsilon}  \tau^{- 1-4\epsilon} 
 \frac{2\pi^{-\epsilon}}{\Gamma(1-\epsilon) }  \frac{1}{2}  \int_0^1    \mathrm{d} x_1 
\mathrm{d} x_2 
\mathrm{d} x_3 \,
 x_1^{3\epsilon}  
\frac{-\epsilon}{x_3^{1+\epsilon}}  
\frac{
\left[ 
2^2 s_{\phi_1}^2 (1-x_3) 
\right]^{-\epsilon}
}{\Big[A_{lk,i}(x_1,c_{\phi_1})\Big]^{1+\epsilon}} \nonumber\\
&& \times \Big[
F^{ij}_l(x_3) + F^{ij}_l(1-x_3) \Big]  \,, 
\end{eqnarray}
where
\begin{equation}
	F^{ij}_l(x_3)=\left[\frac{1}{A_{lk,j}(x_1,c^{\alpha_1}_{\phi_j -  \phi_1})}\right]^{\epsilon}
	\theta\big(A_{lk,i}(x_1,c_{\phi_1})-x_1\big)
\theta\big(A_{lk,j}(x_1,c^{\alpha_1}_{\phi_j -  \phi_1})-x_1\big) \,. 
\end{equation}

\end{itemize}


\begin{thebibliography}{99}


\bibitem{Cieri:2018oms} 
  L.~Cieri, X.~Chen, T.~Gehrmann, E.~W.~N.~Glover and A.~Huss,
  arXiv:1807.11501 [hep-ph].

\bibitem{Dulat:2018bfe}
  F.~Dulat, B.~Mistlberger and A.~Pelloni,
  arXiv:1810.09462 [hep-ph].


\bibitem{Currie:2018fgr} 
  J.~Currie, T.~Gehrmann, E.~W.~N.~Glover, A.~Huss, J.~Niehues and A.~Vogt,
  JHEP {\bf 1805}, 209 (2018)
  doi:10.1007/JHEP05(2018)209
  [arXiv:1803.09973 [hep-ph]].


\bibitem{Currie:2017eqf} 
  J.~Currie, A.~Gehrmann-De Ridder, T.~Gehrmann, E.~W.~N.~Glover, A.~Huss and J.~Pires,
  Phys.\ Rev.\ Lett.\  {\bf 119}, no. 15, 152001 (2017)
  doi:10.1103/PhysRevLett.119.152001
  [arXiv:1705.10271 [hep-ph]].


\bibitem{Boughezal:2015dva} 
  R.~Boughezal, C.~Focke, X.~Liu and F.~Petriello,
  Phys.\ Rev.\ Lett.\  {\bf 115}, no. 6, 062002 (2015)
  doi:10.1103/PhysRevLett.115.062002
  [arXiv:1504.02131 [hep-ph]].



\bibitem{Ridder:2015dxa} 
  A.~Gehrmann-De Ridder, T.~Gehrmann, E.~W.~N.~Glover, A.~Huss and T.~A.~Morgan,
  Phys.\ Rev.\ Lett.\  {\bf 117}, no. 2, 022001 (2016)
  doi:10.1103/PhysRevLett.117.022001
  [arXiv:1507.02850 [hep-ph]].




\bibitem{Boughezal:2015ded} 
  R.~Boughezal, J.~M.~Campbell, R.~K.~Ellis, C.~Focke, W.~T.~Giele, X.~Liu and F.~Petriello,
  Phys.\ Rev.\ Lett.\  {\bf 116}, no. 15, 152001 (2016)
  doi:10.1103/PhysRevLett.116.152001
  [arXiv:1512.01291 [hep-ph]].


\bibitem{Gehrmann-DeRidder:2017mvr} 
  A.~Gehrmann-De Ridder, T.~Gehrmann, E.~W.~N.~Glover, A.~Huss and D.~M.~Walker,
  Phys.\ Rev.\ Lett.\  {\bf 120}, no. 12, 122001 (2018)
  doi:10.1103/PhysRevLett.120.122001
  [arXiv:1712.07543 [hep-ph]].


\bibitem{Boughezal:2015dra} 
  R.~Boughezal, F.~Caola, K.~Melnikov, F.~Petriello and M.~Schulze,
  Phys.\ Rev.\ Lett.\  {\bf 115}, no. 8, 082003 (2015)
  doi:10.1103/PhysRevLett.115.082003
  [arXiv:1504.07922 [hep-ph]].


\bibitem{Chen:2014gva} 
  X.~Chen, T.~Gehrmann, E.~W.~N.~Glover and M.~Jaquier,
  Phys.\ Lett.\ B {\bf 740}, 147 (2015)
  doi:10.1016/j.physletb.2014.11.021
  [arXiv:1408.5325 [hep-ph]].


\bibitem{Boughezal:2015aha} 
  R.~Boughezal, C.~Focke, W.~Giele, X.~Liu and F.~Petriello,
  Phys.\ Lett.\ B {\bf 748}, 5 (2015)
  doi:10.1016/j.physletb.2015.06.055
  [arXiv:1505.03893 [hep-ph]].


\bibitem{Cacciari:2015jma} 
  M.~Cacciari, F.~A.~Dreyer, A.~Karlberg, G.~P.~Salam and G.~Zanderighi,
  Phys.\ Rev.\ Lett.\  {\bf 115}, no. 8, 082002 (2015)
  Erratum: [Phys.\ Rev.\ Lett.\  {\bf 120}, no. 13, 139901 (2018)]
  doi:10.1103/PhysRevLett.115.082002, 10.1103/PhysRevLett.120.139901
  [arXiv:1506.02660 [hep-ph]].



\bibitem{Cruz-Martinez:2018rod} 
  J.~Cruz-Martinez, T.~Gehrmann, E.~W.~N.~Glover and A.~Huss,
  Phys.\ Lett.\ B {\bf 781}, 672 (2018)
  doi:10.1016/j.physletb.2018.04.046
  [arXiv:1802.02445 [hep-ph]].


\bibitem{Brucherseifer:2014ama} 
  M.~Brucherseifer, F.~Caola and K.~Melnikov,
  Phys.\ Lett.\ B {\bf 736}, 58 (2014)
  doi:10.1016/j.physletb.2014.06.075
  [arXiv:1404.7116 [hep-ph]].





\bibitem{Berger:2016oht} 
  E.~L.~Berger, J.~Gao, C.-P.~Yuan and H.~X.~Zhu,
  Phys.\ Rev.\ D {\bf 94}, no. 7, 071501 (2016)
  doi:10.1103/PhysRevD.94.071501
  [arXiv:1606.08463 [hep-ph]].





\bibitem{Czakon:2015owf} 
  M.~Czakon, D.~Heymes and A.~Mitov,
  Phys.\ Rev.\ Lett.\  {\bf 116}, no. 8, 082003 (2016)
  doi:10.1103/PhysRevLett.116.082003
  [arXiv:1511.00549 [hep-ph]].



\bibitem{Catani:2019iny} 
  S.~Catani, S.~Devoto, M.~Grazzini, S.~Kallweit, J.~Mazzitelli and H.~Sargsyan,
  arXiv:1901.04005 [hep-ph].



\bibitem{Abelof:2016pby} 
  G.~Abelof, R.~Boughezal, X.~Liu and F.~Petriello,
  Phys.\ Lett.\ B {\bf 763}, 52 (2016)
  doi:10.1016/j.physletb.2016.10.022
  [arXiv:1607.04921 [hep-ph]].


\bibitem{Currie:2017tpe} 
  J.~Currie, T.~Gehrmann, A.~Huss and J.~Niehues,
  JHEP {\bf 1707}, 018 (2017)
  doi:10.1007/JHEP07(2017)018
  [arXiv:1703.05977 [hep-ph]].




\bibitem{Berger:2016inr} 
  E.~L.~Berger, J.~Gao, C.~S.~Li, Z.~L.~Liu and H.~X.~Zhu,
  Phys.\ Rev.\ Lett.\  {\bf 116}, no. 21, 212002 (2016)
  doi:10.1103/PhysRevLett.116.212002
  [arXiv:1601.05430 [hep-ph]].

\bibitem{Abreu:2018aqd} 
  S.~Abreu, L.~J.~Dixon, E.~Herrmann, B.~Page and M.~Zeng,
  arXiv:1812.08941 [hep-th].


\bibitem{Chicherin:2018yne} 
  D.~Chicherin, J.~M.~Henn, P.~Wasser, T.~Gehrmann, Y.~Zhang and S.~Zoia,
  arXiv:1812.11057 [hep-th].


\bibitem{Chicherin:2018old} 
  D.~Chicherin, T.~Gehrmann, J.~M.~Henn, P.~Wasser, Y.~Zhang and S.~Zoia,
  arXiv:1812.11160 [hep-ph].

\bibitem{Borowka:2018dsa} 
  S.~Borowka, T.~Gehrmann and D.~Hulme,
  JHEP {\bf 1808}, 111 (2018)
  doi:10.1007/JHEP08(2018)111
  [arXiv:1804.06824 [hep-ph]].


\bibitem{Liu:2017jxz} 
  X.~Liu, Y.~Q.~Ma and C.~Y.~Wang,
  Phys.\ Lett.\ B {\bf 779}, 353 (2018)
  doi:10.1016/j.physletb.2018.02.026
  [arXiv:1711.09572 [hep-ph]].

\bibitem{Liu:2018dmc} 
  X.~Liu and Y.~Q.~Ma,
  arXiv:1801.10523 [hep-ph].

\bibitem{Xu:2018eos} 
  X.~Xu and L.~L.~Yang,
  arXiv:1810.12002 [hep-ph].



\bibitem{Boughezal:2015eha} 
  R.~Boughezal, X.~Liu and F.~Petriello,
  Phys.\ Rev.\ D {\bf 91}, no. 9, 094035 (2015)
  doi:10.1103/PhysRevD.91.094035
  [arXiv:1504.02540 [hep-ph]].

\bibitem{Campbell:2017hsw} 
  J.~M.~Campbell, R.~K.~Ellis, R.~Mondini and C.~Williams,
  Eur.\ Phys.\ J.\ C {\bf 78}, no. 3, 234 (2018)
  doi:10.1140/epjc/s10052-018-5732-1
  [arXiv:1711.09984 [hep-ph]].
  
\bibitem{Li:2016tvb} 
  H.~T.~Li and J.~Wang,
  JHEP {\bf 1702}, 002 (2017)
  doi:10.1007/JHEP02(2017)002
  [arXiv:1611.02749 [hep-ph]].
  
\bibitem{Li:2018tsq} 
  H.~T.~Li and J.~Wang,
  Phys.\ Lett.\ B {\bf 784}, 397 (2018)
  doi:10.1016/j.physletb.2018.08.019
  [arXiv:1804.06358 [hep-ph]].

\bibitem{Bell:2018mkk} 
  G.~Bell, B.~Dehnadi, T.~Mohrmann and R.~Rahn,
  PoS LL {\bf 2018}, 044 (2018)
  doi:10.22323/1.303.0044
  [arXiv:1808.07427 [hep-ph]].
  
  
\bibitem{GehrmannDeRidder:2005cm} 
  A.~Gehrmann-De Ridder, T.~Gehrmann and E.~W.~N.~Glover,
  JHEP {\bf 0509}, 056 (2005)
  doi:10.1088/1126-6708/2005/09/056
  [hep-ph/0505111].

  
\bibitem{Czakon:2010td} 
  M.~Czakon,
  Phys.\ Lett.\ B {\bf 693}, 259 (2010)
  doi:10.1016/j.physletb.2010.08.036
  [arXiv:1005.0274 [hep-ph]].

  
\bibitem{Caola:2017dug} 
  F.~Caola, K.~Melnikov and R.~Röntsch,
  Eur.\ Phys.\ J.\ C {\bf 77}, no. 4, 248 (2017)
  doi:10.1140/epjc/s10052-017-4774-0
  [arXiv:1702.01352 [hep-ph]].


\bibitem{Catani:2007vq} 
  S.~Catani and M.~Grazzini,
  Phys.\ Rev.\ Lett.\  {\bf 98}, 222002 (2007)
  doi:10.1103/PhysRevLett.98.222002
  [hep-ph/0703012].

\bibitem{Gao:2012ja} 
  J.~Gao, C.~S.~Li and H.~X.~Zhu,
  Phys.\ Rev.\ Lett.\  {\bf 110}, no. 4, 042001 (2013)
  doi:10.1103/PhysRevLett.110.042001
  [arXiv:1210.2808 [hep-ph]].



\bibitem{Gaunt:2015pea} 
  J.~Gaunt, M.~Stahlhofen, F.~J.~Tackmann and J.~R.~Walsh,
  JHEP {\bf 1509}, 058 (2015)
  doi:10.1007/JHEP09(2015)058
  [arXiv:1505.04794 [hep-ph]].

  
  \bibitem{Stewart:2010tn} 
  I.~W.~Stewart, F.~J.~Tackmann and W.~J.~Waalewijn,
  Phys.\ Rev.\ Lett.\  {\bf 105}, 092002 (2010)
  [arXiv:1004.2489 [hep-ph]].


\bibitem{Bauer:2000ew} 
  C.~W.~Bauer, S.~Fleming and M.~E.~Luke,
  Phys.\ Rev.\ D {\bf 63}, 014006 (2000)
  [hep-ph/0005275].

\bibitem{Bauer:2000yr}
C.~W. Bauer, S.~Fleming, D.~Pirjol, and I.~W. Stewart,
\newblock Phys. Rev. {\bf D63}, 114020 (2001), hep-ph/0011336.

\bibitem{Bauer:2001ct} 
  C.~W.~Bauer and I.~W.~Stewart,
  Phys.\ Lett.\ B {\bf 516}, 134 (2001)
  [hep-ph/0107001].

\bibitem{Bauer:2001yt}
C.~W. Bauer, D.~Pirjol, and I.~W. Stewart,
\newblock Phys. Rev. {\bf D65}, 054022 (2002), hep-ph/0109045.

\bibitem{Bauer:2002nz}
C.~W. Bauer, S.~Fleming, D.~Pirjol, I.~Z. Rothstein, and I.~W. Stewart,
\newblock Phys. Rev. {\bf D66}, 014017 (2002), hep-ph/0202088.


\bibitem{Moult:2016fqy} 
  I.~Moult, L.~Rothen, I.~W.~Stewart, F.~J.~Tackmann and H.~X.~Zhu,
  Phys.\ Rev.\ D {\bf 95}, no. 7, 074023 (2017)
  doi:10.1103/PhysRevD.95.074023
  [arXiv:1612.00450 [hep-ph]].


\bibitem{Boughezal:2016zws} 
  R.~Boughezal, X.~Liu and F.~Petriello,
  JHEP {\bf 1703}, 160 (2017)
  doi:10.1007/JHEP03(2017)160
  [arXiv:1612.02911 [hep-ph]].
  
\bibitem{Moult:2017jsg} 
  I.~Moult, L.~Rothen, I.~W.~Stewart, F.~J.~Tackmann and H.~X.~Zhu,
  Phys.\ Rev.\ D {\bf 97}, no. 1, 014013 (2018)
  doi:10.1103/PhysRevD.97.014013
  [arXiv:1710.03227 [hep-ph]].

\bibitem{Boughezal:2018mvf} 
  R.~Boughezal, A.~Isgrò and F.~Petriello,
  Phys.\ Rev.\ D {\bf 97}, no. 7, 076006 (2018)
  doi:10.1103/PhysRevD.97.076006
  [arXiv:1802.00456 [hep-ph]].


\bibitem{Ebert:2018lzn} 
  M.~A.~Ebert, I.~Moult, I.~W.~Stewart, F.~J.~Tackmann, G.~Vita and H.~X.~Zhu,
  JHEP {\bf 1812}, 084 (2018)
  doi:10.1007/JHEP12(2018)084
  [arXiv:1807.10764 [hep-ph]].



\bibitem{Gaunt:2014xga} 
  J.~R.~Gaunt, M.~Stahlhofen and F.~J.~Tackmann,
  JHEP {\bf 1404}, 113 (2014)
  [arXiv:1401.5478 [hep-ph]].
 
\bibitem{Gaunt:2014cfa} 
  J.~Gaunt, M.~Stahlhofen and F.~J.~Tackmann,
  JHEP {\bf 1408}, 020 (2014)
  [arXiv:1405.1044 [hep-ph]].

\bibitem{Becher:2006qw} 
  T.~Becher and M.~Neubert,
  Phys.\ Lett.\ B {\bf 637}, 251 (2006)
  [hep-ph/0603140].

\bibitem{Becher:2010pd} 
  T.~Becher and G.~Bell,
  Phys.\ Lett.\ B {\bf 695}, 252 (2011)
  [arXiv:1008.1936 [hep-ph]].
  
  
  
\bibitem{Bruser:2018rad} 
  R.~Brüser, Z.~L.~Liu and M.~Stahlhofen,
  Phys.\ Rev.\ Lett.\  {\bf 121}, no. 7, 072003 (2018)
  doi:10.1103/PhysRevLett.121.072003
  [arXiv:1804.09722 [hep-ph]].

  
\bibitem{Banerjee:2018ozf} 
  P.~Banerjee, P.~K.~Dhani and V.~Ravindran,
  Phys.\ Rev.\ D {\bf 98}, no. 9, 094016 (2018)
  doi:10.1103/PhysRevD.98.094016
  [arXiv:1805.02637 [hep-ph]].

  
\bibitem{Melnikov:2018jxb} 
  K.~Melnikov, R.~Rietkerk, L.~Tancredi and C.~Wever,
  arXiv:1809.06300 [hep-ph].


\bibitem{Boughezal:2017tdd} 
  R.~Boughezal, F.~Petriello, U.~Schubert and H.~Xing,
  Phys.\ Rev.\ D {\bf 96}, no. 3, 034001 (2017)
  doi:10.1103/PhysRevD.96.034001
  [arXiv:1704.05457 [hep-ph]].



  
  
  
  
\bibitem{Catani:1999ss} 
  S.~Catani and M.~Grazzini,
  Nucl.\ Phys.\ B {\bf 570}, 287 (2000)
  [hep-ph/9908523].

\bibitem{Catani:2000pi} 
  S.~Catani and M.~Grazzini,
  Nucl.\ Phys.\ B {\bf 591}, 435 (2000)
  [hep-ph/0007142].

\bibitem{Alioli:2013hqa} 
  S.~Alioli, C.~W.~Bauer, C.~Berggren, F.~J.~Tackmann, J.~R.~Walsh and S.~Zuberi,
  JHEP {\bf 1406}, 089 (2014)
  doi:10.1007/JHEP06(2014)089
  [arXiv:1311.0286 [hep-ph]].


\end{thebibliography}
\end{document}